\documentclass[aps,prl,twocolumn,groupedaddress,showpacs]{revtex4-1}
\usepackage{graphicx}
\usepackage{amsmath}
\begin{document}

%\preprint{}

\title{Retrieving the Size of Deep-subwavelength Objects via \\Tunable Optical Spin-Orbit Coupling }


\author{Zheng Xi}
\email[z.xi@tudelft.nl]{}
\affiliation{Optica Reseach Group, Delft University of Technology}
\author{H.P. Urbach}
%\thanks{}
\affiliation{Optica Reseach Group, Delft University of Technology}


\date{\today}

\begin{abstract}
We propose a scheme to retrieve the size parameters of a nano-particle on a glass substrate at a scale much smaller than the wavelength. This is achieved by illuminating the particle using two plane waves to create rich and non-trivial local polarization distributions, and observing the far-field scattering pattern into the substrate. A simple dipole model which exploits tunneling effect of evanescent field into regions beyond the critical angle, as well as directional scattering due to spin-orbit coupling is developed, to relate the particle's shape, size and position to the far-field scattering with remarkable sensitivity. Our method brings about a far-field super-resolution imaging scheme based on the interaction of vectorial light with nanoparticles.
\end{abstract}
%\keywords{}
\maketitle
%\section{}
The great interest in nanotechnology demands a simple, non-invasive and far-field optical technique that is able to provide precise information about the shape, size and morphology of individual nanoparticles, in order to monitor and use their size-dependent properties. The problem in a standard optical microscopy is the diffraction barrier which predicts that for particles of deep-subwavelength dimensions, although they may vary in size, would appear nearly as the same size of about $\lambda_0/{2NA}$ in the far-field (point spread function), where $\lambda_0$ is the illumination wavelength, NA is the numerical aperture of the lens. While breakthrough in super-resolution fluorescent microscopy makes use of non-linear optical effects to achieve super-resolution via isolation and localization of single fluorescent molecules\cite{RN1,RN2,RN3,RN4}, this approach cannot be applied directly to the case of non-fluorescent nanoparticles. 

Light field containing inhomogeneous polarization distributions has attracted great attention\cite{RN5,RN6,RN7,RN8,RN9,RN10}. Of particular interest is the local transverse spin in  optical fields\cite{RN9,RN11,RN12}. In analogy to the spin-hall effect of electrons, photons of different spins are found to couple to different directions. This intrinsic spin-dependent-propagation not only intrigues numerous studies in classical optics\cite{RN13,RN14,RN15,RN16,RN17,RN18,RN19} but also is considered as a fundamental building block of future chiral quantum networks\cite{RN20,RN21,RN22,RN23}. Most of previous research focuses on the forward problem: encoding the spin information into different directions of propagation\cite{RN9,RN24,RN25,RN26}. In this Letter, we try to tackle the inverse problem: retrieving the subwavelength dimension of nanoparticles based on tunable spin-dependent-scattering into the far-field. In particular, we  consider two cases of nanoparticles on a glass substrate with ${{n}_{glass}}=1.5$: a nanowire of subwavelength elliptical cross-section characterized by two independent axes (2D case), and an ellipsoidal nanoparticle characterized by three independent axes (3D case). The material of the particle is chosen to be silicon with refractive index $n=3.5$ to match the standard silicon-on-insulator technology. By illuminating the nanoparticle with an interference field and applying a simple dipole model, it is found that the exponential decay of the scattered power beyond the critical angle can be used to retrieve the particle's height, while the local spin-driven directional scattering can be used to retrieve the transverse-to-vertical aspect ratio. This inversion scheme directly links the far-field scattering with deep-subwavelength information of the nanoparticle without sophisticated and time-consuming optimization algorithms, opening new possibilities for the future optical metrology methods exploiting the local interaction of vectorial light and matter.  
\begin{figure}
\includegraphics[width=\linewidth]{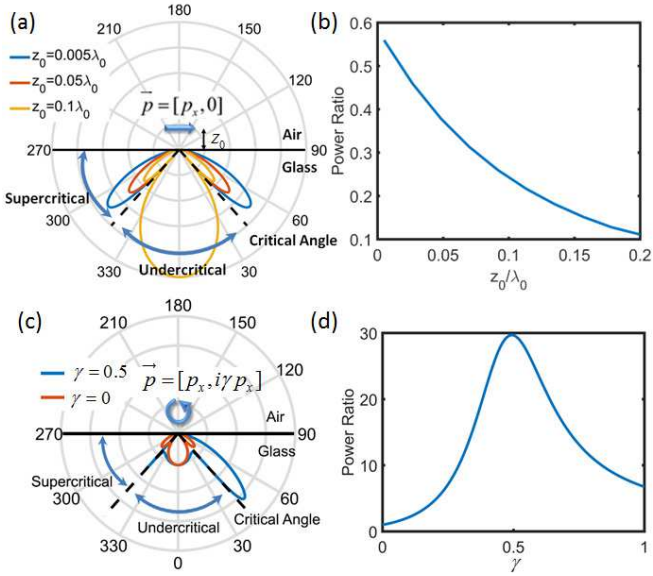}
\caption{(a) Far-field radiation pattern into the glass substrate when a horizontal dipole ${p}_{x}$ is placed at different distances ${{z}_{0}}$ from the interface. The radiation pattern in UAR is the same for all $z_0$. (b) Power ratio between the SAR and the UAR as the horizontal dipole is moved away from the interface. (c) Far-field radiation pattern into the glass substrate for a spinning dipole. (d) Power ratio between the RSAR and the LSAR for different $\gamma$ for a general spinning dipole located at ${{z}_{0}}=0.005{{\lambda }_{0}}$.}
\label{fig:false-color}
\end{figure}

To start with, we consider the 2D case of the emission of a general dipole $\overrightarrow{p}=[{{p}_{x}},{{p}_{z}}]$ above a glass substrate with surface normal parrallel to the z-axis. Since the scattering of a Rayleigh particle can be well described by this model, this serves as the theoretical foundation of the scheme.

The dipole is put in air with a distance ${{z}_{0}}$ from the interface. The magnetic field radiated into the substrate can be described by the angular spectrum method\cite{RN27}:
\begin{equation}
{{H}_{y}}({{k}_{x}})=\frac{i{{k}_{zg}}}{8{{\pi }^{2}}}{{t}_{p}}({{p}_{x}}+\frac{{{k}_{x}}}{{{k}_{za}}}{{p}_{z}}){{e}^{i{{k}_{za}}{{z}_{0}}}}
\label{}
\end{equation}
where ${{k}_{zg}}=\sqrt{n_{g}^{2}k_{0}^{2}-k_{x}^{2}}$,  ${{k}_{za}}=\sqrt{k_{0}^{2}-k_{x}^{2}}$ are the vertical wave vectors in glass and in air, ${{k}_{0}}$ is the wavenumber in air, ${{n}_{g}}$ is the refractive index of glass, and ${{t}_{p}}$ is the Fresnel coefficient. The term ${{e}^{i{{k}_{za}}{{z}_{0}}}}$ in Eq. (1) separates propagating (${{k}_{x}}\le {{k}_{0}}$, ${{k}_{za}}$ is real)  and evanescent waves (${{k}_{x}}>{{k}_{0}}$, ${{k}_{za}}$ is imaginary) into two regions defined by the critical angle: the undercritical angle region (UAR) and supercritical angle region (SAR) (Fig. 1(a)). We first consider the case where ${{p}_{z}}=0$, i.e. an x-polarized dipole, and change the dipole-to-interface distance ${{z}_{0}}$ from $0.005{{\lambda }_{0}}$ to $0.1{{\lambda }_{0}}$. The power in the SAR comes from evanescent waves and thus increases exponentially as ${{e}^{-|{{k}_{za}}|{{z}_{0}}}}$ when ${{z}_{0}}$ decreases. In contrast, the power within the UAR originates from propagating waves and therefore the field only gains an extra phase term ${{e}^{i{{k}_{za}}{{z}_{0}}}}$ leaving the total power in this region unchanged. The power ratio of the scattered field in the UAR and the SAR can be calculated as a function of ${{z}_{0}}$ and the result is shown in Fig. 1(b). The exponential decay of the power ratio can be used to extract the vertical position of the dipole, even at deep subwavelength scale\cite{RN28,RN29,RN30}. Next we consider a more general case: a spinning dipole with ${{p}_{z}}=i\gamma {{p}_{x}}$ with varying $\gamma$ from 0 to 1 (Fig. 1(c) and (d)). The additional spin introduced $\pi /2$ phase difference breaks the symmetry depending on the handness of the dipole\cite{RN9}.  This in turn leads to the uneven distribution of the angular spectrum for ${{k}_{x}}>{{k}_{0}}$, which can then be translated into asymmetric propagation in the SAR (Fig. 1(c)). This spin-dependent directional emission is very sensitive to the value of $\gamma$. In Fig. 1(d), we plot the integrated power ratio into the right supercritical angle region (RSAR) and left supercritical angle region (LSAR) against different $\gamma$ at $z_0=0.005\lambda_0$. An optimum ${{\gamma }_{opt}}=0.5$ that yields the largest asymmetry is clearly visible.

From the above discussions, we have seen for a simple dipole model that the far-field changes significantly with the proper choice of the complex dipole moment. Precise information about the dipole’s vertical position ${{z}_{0}}$ as well as the optimum ratio ${{\gamma }_{opt}}$ between different dipole moments can be extracted from far-field measurements. Replacing the dipole  by a Rayleigh particle, we wonder if it is possible to develop a quantitative approach to retrieve subwavelength information about the particle based on the same principle.

We first consider the 2D case: a silicon nanowire with an elliptical cross-section characterized by two axis of lengths $a$ and $b$ in the x and z directions respectively. The nanowire is sitting on a glass substrate. This geometry is widely used in nano-waveguide applications. The cross-section profile plays an important role in the determination of the overall performance of the waveguide. The aim here is to retrieve the length of the two axes $a$ and $b$ from the far-field measurements. For this purpose, the illumination field is designed to be two p-polarized plane waves forming an interference pattern along the air-glass interface shown in Fig. 2(a). To simplify the discussion, the incident angle is chosen to be at the Brewster angle $\theta_p$ for both beams. The illumination field above the interface can be calculated analytically:
\begin{equation}
\begin{split}
  & {{E}_{x}}=2{{E}_{0}}\cos {{\theta }_{p}}\cos ({{k}_{x}}x), \\ 
  & {{E}_{z}}=2i{{E}_{0}}\sin {{\theta }_{p}}\sin ({{k}_{x}}x), 
\end{split}
\end{equation}
where ${{E}_{0}}$ is the amplitude of the incident beam, ${{k}_{x}}$ is the transverse wave vector component in air.  Interestingly, even for this seemingly simple field, it contains rich local polarization topology distributions along the x axis\cite{RN18}. Starting from $x=0$, the local polarization state changes from a singular L-line (linear polarization) gradually to C-line (circular polarization) of different handness on each side as shown schematically in Fig. 2(a). It is also worth noting the role of polarization singularity here. If a detection scheme can be developed such that it responses only to a certain polarization state, the slight deviation from that polarization singularity would cause significantly changed observables, making the scheme very sensitive\cite{RN13,RN19}.  

To utilize this polarization singularity to obtain a very sensitive detection, we consider a nanowire of a certain shape parameters $a$ and $b$ placed inside the field and calculate the power ratio in RSAR and LSAR at different x for different shapes (Fig. 2(b)). When the local polarization state at a certain displacement ${{x}_{0}}$ matches the shape of the cross-section, resulting in optimum $\gamma_{opt}$ between induced complex dipole moments, the power in the SAR becomes highly asymmetric (peaks in Fig. 2(b)). As a result, different shape information is encoded in the displacement measurement along x.

\begin{figure}
\includegraphics[width=\linewidth]{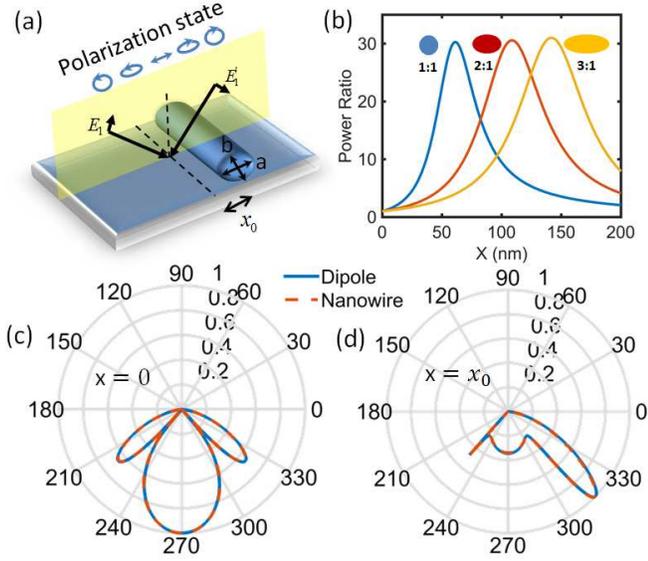}
\caption{(a) Geometry of the considered 2D system. A nanowire is placed on top of a glass substrate. Two p-polarized plane waves are incident at the Brewster angle on the nanowire, forming an interference pattern. The nanowire is moved along the x axis. The polarization state along x changes from linear to circular of different handness as schematically shown by the blue arrows and circles. (b) Power ratio in RSAF and LSAF region as the nanowire is moving along x. The insets show different shapes of the cross-sections with aspect ratio $a:b=1:1$(blue), $2:1$(red) and $3:1$(yellow) when $b=10nm$. (c) Far-field scattering (red dashed curve) pattern of a nanowire placed at $x=0$ into the substrate. The scattering pattern can be well described by an equivalent horizontal dipole placed at the center of the nanowire (blue curve). (d) Same as Fig. 2(c) except the nanowire is placed at $x=x_0$ and can be approximated by a spinning dipole with $\gamma_{opt}$. }
\label{fig:false-color}
\end{figure}

To decode the relation between the displacement and the shape parameters, an analytical model can be derived. Within the Born approximation and assuming $a,b<<\lambda$, the two induced dipoles inside the nanowire can be written as:
\begin{equation}
\begin{split}
  & {{p}_{x}}={{\alpha }_{x}}{{E}_{x}}, \\ 
  & {{p}_{z}}={{\alpha }_{z}}{{E}_{z}}, 
\end{split}
\end{equation}
with the polarizability related to the geometry parameters\cite{RN31}:
\begin{equation}
{{\alpha }_{i}}=V\frac{{{\epsilon }_{r}}-1}{1+{{L}_{i}}({{\epsilon }_{r}}-1)}\quad i=x,z,
\end{equation}
where $\epsilon_r$ is the relative permittivity of the nanowire, ${{L}_{x}}=\frac{a}{a+b},\,\ {{L}_{z}}=\frac{b}{a+b}$ corresponds to the geometry of the cross-section and $V={\pi^2}{ab}$ .  

Combing Eq. (3) and (4), the ratio between the induced dipole components can be calculated as:
\begin{equation}
\frac{{{p}_{x}}}{{{p}_{z}}}=\frac{s+{{\epsilon }_{r}}}{1+s{{\epsilon }_{r}}}\frac{{{E}_{x}}}{{{E}_{z}}}
\end{equation}
with $s=a/b$ is the aspect ratio of the two axes, which relates the geometry of the two axes with the induced dipoles.

When the background field is taken to be as in Eq. (2), the above expression can be written as:
\begin{equation}
\frac{{{p}_{x}}}{{{p}_{z}}}=i\cot {{\theta }_{p}}\cot ({{k}_{x}}x)\frac{s+{{\epsilon }_{r}}}{1+s{{\epsilon }_{r}}}
\end{equation}
hence, a general dipole with a changing ellipticity depending on the location x is induced. Consider the range $x\in (-\frac{\pi }{2{{k}_{x}}},\frac{\pi }{2{{k}_{x}}})$. Only at $x=0$, the illumination field is linearly x-polarized (L-line polarization singularity), and the nanowire can be approximated by a linearly x-polarized dipole situated at the center of the nanowire (Fig. 2(c)). The scattered field into the substrate in this case is symmetric. The vertical position of the induced dipole corresponds to the height $b/2$ of the nanowire, and can be extracted from the power ratio between UAR and SAR. Because the local polarization state changes continuously along x, at a certain displacement ${{x}_{0}}$ from the origin, the nanowire becomes a spinning dipole with optimized ${{\gamma }_{opt}}$ (Fig. 2(d)). By substituting ${{\gamma }_{opt}}$ into Eq. (6) one can get the solution for the aspect ratio $s$:
\begin{equation} 
s=\frac{{{\epsilon }_{r}}-A}{A{{\epsilon }_{r}}-1}
\end{equation}
where $A=\tan ({{\theta }_{p}})\tan({{k}_{x}}{{x}_{0}})/|{{\gamma }_{opt}}|$. This simple expression relates the aspect ratio $s$ directly to the optimum displacement ${{x}_{0}}$ which yields the largest asymmetry in the far-field. From the aspect ratio $s$ and the length of the vertical axis $b$, the transverse dimension $a$ can be retrieved. This far-field measurement approach gives exact information of the particle’s shape (aspect ratio $s$), size (length of the two axes $a$ and $b$) and location ${{x}_{0}}$ at the same time. 

In Fig. 3, several geometries with different parameters are tested. The illumination wavelength is chosen to be $1000nm$ for all cases. Fig. 3(a) shows the retrieval of the height of a silicon nanowire with a circular cross-section. The only unknown here is the length of the vertical axis $b$. For the retrieval of  the aspect ratio $s$, $b$ is fixed at $10nm$. The results are shown in Fig. 3(b). In both cases, the retrieved results match the original design quite well.
		
\begin{figure}
\includegraphics[width=\linewidth]{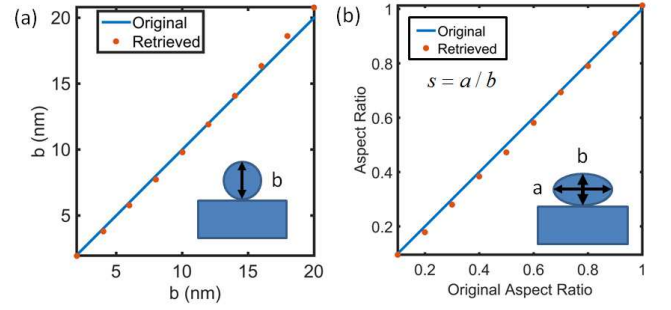}
\caption{(a) Retrieval of the height $b$ of silicon nanowires of circular cross-section from the power ratio in the UAR and SAR. (b) Retrieval of the aspect ratio from the power ratio of the RSAR and LSAR. The height $b$ is fixed as $10nm$.}
\label{fig:false-color}
\end{figure}

The above discussions are for 2D nanowaveguide applications, the proposed scheme can be applied to a general 3D case as well. The 3D particle can be described by an ellipsoidal nanorod with three axes $a$, $b$ and $c$ shown in Fig. 4(a).  We assume the orientation of the nanorod can be determined by polarization analysis as done in previous works\cite{RN32,RN33,RN34}. The main goal here is to retrieve $a$, $b$ and $c$.  

The vertical dimension $c$ can be retrieved as before by looking at the power ratio in SAR and UAR. For $a$ and $b$, two successive illuminations are applied with p-polarized plane waves along YZ-plane and XZ-plane respectively. The induced dipoles at the optimum displacement ${{x}_{0}}$ and ${{y}_{0}}$ can be calculated as:
\begin{equation}
\begin{split}
  & \frac{{{p}_{x}}}{{{p}_{z}}}=i\frac{{{\alpha }_{x}}}{{{\alpha }_{z}}}\cot {{\theta }_{p}}\cot {{k}_{x}}{{x}_{0}}=\frac{1}{i|{{\gamma }_{opt}}|}, \\ 
 & \frac{{{p}_{y}}}{{{p}_{z}}}=i\frac{{{\alpha }_{y}}}{{{\alpha }_{z}}}\cot {{\theta }_{p}}\cot {{k}_{y}}{{y}_{0}}=\frac{1}{i|{{\gamma }_{opt}}|}, \\ 
\end{split}
\end{equation}
with ${{\alpha }_{x}}$, ${{\alpha }_{y}}$ and $\alpha_z$ determined analogously to Eq. (4)\cite{RN31}. By substituting Eq. (4) into Eq. (8), we get:
\begin{equation}
\begin{split}
 & A({{\epsilon }_{r}}-1){{L}_{1}}-({{\epsilon }_{r}}-1){{L}_{3}}=1-A, \\ 
 & B({{\epsilon }_{r}}-1){{L}_{2}}-({{\epsilon }_{r}}-1){{L}_{3}}=1-B, \\ 
\end{split}
\end{equation}
in which ${{\epsilon }_{r}}$ is the permittivity of the nanorod, $A=\tan{{\theta }_{p}}\tan ({{k}_{y}}{{y}_{0}})/|{{\gamma }_{opt}}|$ and $B=\tan{{\theta }_{p}}\tan({{k}_{x}}{{x}_{0}})/|{{\gamma }_{opt}}|$ . Eq. (8) together with the requirement that\cite{RN31}:
\begin{equation}
{{L}_{1}}+{{L}_{2}}+{{L}_{3}}=1\\ 
\end{equation}
can be used to solve ${{L}_{1}}$, ${{L}_{2}}$ and ${{L}_{3}}$ based on the measured values of ${{x}_{0}}$ and ${{y}_{0}}$.
To further relate these values to the geometrical parameters $a$, $b$ and $c$, we compare the ${{L}_{1}}$, ${{L}_{2}}$ and ${{L}_{3}}$ with the theoretically calculated ones according to\cite{RN31}:
\begin{equation}
\begin{split}
& {{L}_{i}}=\int\limits_{0}^{\infty }{\frac{{{l}_{i}}{{l}_{j}}{{l}_{k}}dm}{2{{(m+l_{i}^{2})}^{3/2}}{{(m+l_{j}^{2})}^{1/2}}{(m+l_{k}^{2})}^{1/2}}}\\ 
\end{split}
\end{equation}
in which $l_{i,j,k}=a/2$, $b/2$ and $c/2$ respectively by varying $a$, $b$ and $c$ to get the best match.
 
As one example, we consider a silicon nanorod sitting on a glass substrate. The wavelength is $1000nm$. The three axes are chosen as $a=30nm$, $b=20nm$ and $c=10nm$.  The calculated power ratios in the SAR  are plotted in the inset of Fig. 4(a). From this, the optimum ${{x}_{0}}=114nm$ and ${{y}_{0}}=149nm$ can be extracted. Substituting these values in Eqs.(9)-(11), the information of the three axes are retrieved and the results are shown in the bottom of Fig. 4(a).

\begin{figure}
\includegraphics[width=\linewidth]{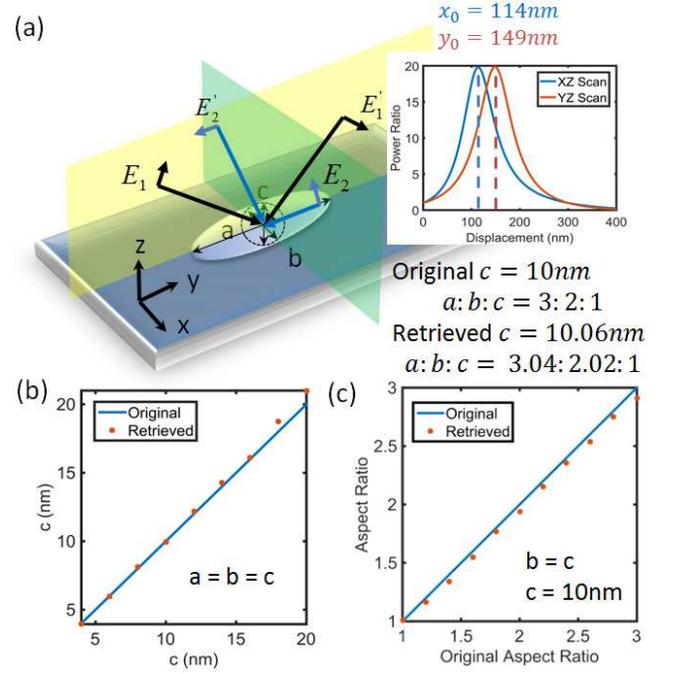}
\caption{(a) Geometry of the considered 3D system. A nanorod is placed on a glass substrate. Two successive illuminations in the YZ (yellow) plane and XZ (green) plane are used to illuminate the nanorod. The calculated optimum displacements ${{x}_{0}}$ and ${{y}_{0}}$ are shown in the top-right inset for a nanorod of dimenions $a=30nm$, $b=20nm$ and $c=10nm$. Bottom right is the retrieved parameters. (b) Retrieval of the radius of a nanosphere. (c) Retrieval of the aspect ratio $a/c$ of different nanorods with fixed height $c=10nm$.} 
\label{fig:false-color}
\end{figure}
For most situations, the nanorod can be approximated as a prolate spheroid ($a>b=c$). We have also investigated this case and the results are shown in Fig. 4(b) and (c). The retrieval of the length of the vertical axis $c$ is shown in Fig. 4(b). Nanospheres of changing radius $c$ are considered as examples. In Fig. 4(c), the vertical axis is fixed at $c=10nm$ and the aspect ratio $a/c$ is retrieved. 

Finally, it is also worth mentioning that the displacement of the particle can be equivalently treated as changing the phase of one of the incident plane wave. Because the control over phase can be very accurate, it may gain additional values from experimental point of view. 

In conclusion, we have proposed a method to retrieve precise information about the shape, size and location of a particle at a deep subwavelength scale. The upper limit of the particle's size is imposed by the condition that the particle is much smaller than the wavelength such that quasi-static approximation for the polarizability holds. The hard problem of achieving subwavelength information is converted into the measurement of the symmetry of the scattering pattern at different locations. The method utilizes the full vectorial interaction of light and particle, in particular, the photonic spin-orbit interaction, and it also serves as a guideline for the development of ultrasensitive displacement sensors by shaping nanoparticles\cite{RN10,RN19,RN35,RN36}. Additionally, because the model retrieves the complex dipole properties, it may also find interesting applications to “image” complex dipole moments of a single molecule\cite{RN37}. This spin-based retrieval method can have important applications in ultrahigh resolution nanometrology and can shed new light on super-resolution techniques involving the interaction of vectorial light and matter.

%
% ****** End of file apstemplate.tex ******

\end{document}